%% file: connectivity.tex
\documentclass[conference]{IEEEtran}

\makeatletter

\usepackage{blindtext}
\usepackage{eso-pic}
\usepackage{cite}
\usepackage{amsmath,amssymb,amsfonts}
\usepackage{algorithmic}
\usepackage{graphicx}
\usepackage{textcomp}
\usepackage{xcolor}
\usepackage{array,multirow}
\usepackage{graphicx}
\usepackage{tabularx}
\usepackage{tabulary}
\usepackage{latexsym}
\usepackage{textgreek}
\usepackage{layout}
\usepackage{rotating}
\usepackage{amsmath,amsfonts,amssymb}
\usepackage{bm}
\usepackage{dirtytalk}
\usepackage{euscript}
\usepackage{algorithmic}
\usepackage{amssymb, amsmath}
\usepackage[numbers, square, comma, sort&compress]{natbib}
\usepackage[ruled,lined]{algorithm2e}
\usepackage{epstopdf}
\usepackage{gensymb}
\usepackage{subcaption}
\usepackage{siunitx}
\usepackage{acronym}
\usepackage{mathtools}
\usepackage[verbose]{wrapfig}
\usepackage{multirow}
\usepackage[normalem]{ulem}
\def\BibTeX{{\rm B\kern-.05em{\sc i\kern-.025em b}\kern-.08em
    T\kern-.1667em\lower.7ex\hbox{E}\kern-.125emX}}
    
\usepackage{eso-pic}

\usepackage [autostyle, english = american]{csquotes}
\MakeOuterQuote{"}
\usepackage{caption} % to use so that the caption and the table title appears in the same line.
   \usepackage[belowskip=2pt, aboveskip=-13pt] {caption}
\DeclarePairedDelimiter{\ceil}{\lceil}{\rceil}

\newcommand{\eq}[1]{(\ref{#1})}

\acrodef{API}[API]{Application Programmable Interfaces}
\acrodef{ARPU}[ARPU]{Average Revenue per User}
\acrodef{BS}[BS]{Base Station}
\acrodef{CNN}[CNN]{Convolutional Neural Network}
\acrodef{CPU}[CPU]{Central Processing Unit}
\acrodef{DRC-RS}[DRC-RS]{Dynamic Resource Controller for Remote/Rural Sites}
\acrodef{EB}[EB]{Energy Buffer}
\acrodef{EH}[EH]{Energy Harvesting}
\acrodef{ES}[ES]{Energy Saving}
\acrodef{EM}[EM]{Energy Manager}
\acrodef{GENM}[GENM]{Green-based Edge Network Management}
\acrodef{EPC}[EPC]{Evolved Packet Core}
\acrodef{ETSI}[ETSI]{European Telecommunications Standards Institute}
\acrodef{GP}[GP]{Geometric Programming}
\acrodef{ITS}[ITS] {Intelligent Transport System}
\acrodef{LOC}[LOC]{User Location Services} 
\acrodef{LLC}[LLC]{Limited Lookahead Control}
\acrodef{LS}[LS]{Location Service}
\acrodef{LSTM}[LSTM]{Long Short-Term Memory}
\acrodef{MEC}[MEC]{Multi-access Edge Computing}
\acrodef{ML}[ML]{Machine Learning}
\acrodef{MN}[MN]{Mobile Network}
\acrodef{TIM}[TIM]{Telecom Italia Mobile}

\acrodef{NEF}[NEF]{Network Exposure Function}
\acrodef{NOES}[NOES]{NO Energy Saving}
\acrodef{NFV}[NFV]{Network Function Virtualization}
\acrodef{NIC}[NIC]{Network Interface Card}
\acrodef{QoS}[QoS]{Quality of Service}
\acrodef{RNN}[RNN]{Recurrent Neural Network}
\acrodef{RAN}[RAN]{Radio Access Network}
\acrodef{RMSE}[RMSE]{Root Mean Square Error}
\acrodef{RNN}[RNN]{Recurrent Neural Network}
\acrodef{SDN}[SDN]{Software Defined Networking}
\acrodef{UE}[UE]{User Equipment}
\acrodef{VM}[VM] {Virtual Machine}
\acrodef{VNF}[VNF]{Virtualized Network Function}

\begin{document}
\title{\vspace*{1cm} Connect Everywhere: Wireless Connectivity in Protected Areas\\
}

\author{\IEEEauthorblockN{Thembelihle Dlamini, Mengistu Abera Mulatu}
\IEEEauthorblockA{\textit{Department of Electrical and Electronic Engineering} \\
\textit{University of Eswatini}\\
Manzini, Eswatini\\
\{tldlamini, mamulatu\}@uniswa.sz}
\and
\IEEEauthorblockN{Sifiso Vilakati}
\IEEEauthorblockA{\textit{Department of Statistics and Demography} \\
\textit{University of Eswatini}\\
Manzini, Eswatini \\
svilakati@uniswa.sz}
}
\maketitle
%\conf{\textit{  Proc. of the International Conference on Electrical, Computer and Energy Technologies (ICECET) \\ 
%9-10 December 2021, Cape Town-South Africa}}
\begin{abstract}
Mobile connectivity has become more important, especially for visitors to parks and protected areas. However, governments' policies prohibit cable wiring in these areas in order to preserve the beauty and protect the historical interest of the landscape. As a result, in order to provide emergency and infotainment services, mobile connectivity is of primary importance in areas such as national parks and historical sites. Through observed practices from other countries, mobile network operators and other licensed service providers can cooperate with the administrators of such protected areas in order to provide mobile connectivity without disturbing the environment. However, the most pervasive problem is the high energy consumption of the wireless systems and it becomes expensive to power them using the electricity grid. One attractive solution is to make use of green energy to power the communication systems and then share the base station (BS) infrastructure that is \mbox{co-located} with the edge server, an entity responsible for computing the offloaded \mbox{delay-sensitive} workloads. To alleviate this problem, this paper offers a resource management solution that seeks to minimize the energy consumption per communication site through (i) BS infrastructure and resource sharing, and (ii) assisted (\mbox{peer-to-peer}) computation offloading for the \mbox{energy-constrained} communication sites, within a protected area. Using this resource management strategy guarantees a Quality of Service (QoS). The performance evaluation conducted through simulations validates our analysis as the prediction variations observed shows greater accuracy between the harvested energy and the traffic load. Towards energy savings, the proposed algorithm achieves a $52 \%$
energy savings when compared with the $46 \%$ obtained by our benchmark algorithm. The energy savings that can be achieved decreases as the QoS is prioritized, within each communication site, and  when the number of active computing resources increases.
\end{abstract}
%\copyrightnotice{XXX-X-XXXX-XXXX-X/XX/\$XX.00 ©20XX IEEE}

\section{Introduction}
Communication has become the most critical aspect of human life. Hence, mobile connectivity is crucial in offering the required communication for humans wherever they are. \mbox{Multi-connectivity} solutions have been suggested for wireless users in different kinds of environments, however, due to certain governments' policies, the same cannot be said for protected areas~\cite{regulations}. Most governements do not approve cable wiring in such protected areas in order to preserve their beauty. Thus, in order to guarantee mobile connectivity in protected areas, i.e., the provision of emergency services and the deployment of energy \mbox{self-sufficient} communication sites, future \acp{MN} are expected to leverage the integration of \ac{MEC} and \ac{EH} BS, i.e., the \acp{BS} are empowered with computing capabilities and also equipped with \ac{EH} equipments~\cite{edge_controller}. The use of \ac{EH} systems motivates the need for energizing edge systems with green energy in order to extend network coverage to protected areas, and also minimizing the carbon footprint~\cite{energymanagershow}. The \mbox{renewable-powered} \acp{BS} are more suitable in national parks and historical sites, when considering environmental impact, as most governments do not approve cable wiring in order to preserve the landscape~\cite{regulations}.\\
Despite of the potential presented by the integration of \ac{MEC} and \ac{EH} \ac{BS}, the challenges of both resource provisioning and energy consumption still come up under this new paradigm of softwarized \acp{MN}. In addition, the variation of the haversted energy per communication site brings about the notion of \textit{assisted computation offloading}, i.e., one \ac{BS} accept the offloaded \mbox{delay-sensitive} workload and then cooperate with neighboring BSs that have enough green energy and available computing resources (\acp{VM} or containers) in order for them to compute the task and return the result to it. The task offloading technique helps to avoid large computation latency at overloaded BSs or \mbox{energy-deficient} BSs that might degrade the \ac{QoS} to end users~\cite{chen2018computation}. Therefore, in order to address the connectivity issue, this paper formulates an optimization problem to manage a shared \ac{BS} system network (the BS infrastructure and its \mbox{co-located} computing platform (MEC server)) that is deployed within a national park. The primary objective of this paper is to minimize energy consumption through infrastructure sharing and resource management procedures, and assisted computation offloading for the \mbox{energy-constrained} BS sites. Here, the offloaded workloads are jointly allocated over the BS sites and the computing platform resources, in an \mbox{energy-efficient} manner with a guarantee of \ac{QoS}. The formulated problem is solved by exploiting predictions of the BS traffic load, harvested energy, and computation load via a \ac{LLC} approach, obtaining energy savings higher than $50 \%$ with respect to myopic allocation strategies, i.e., with respect to communication site management schemes that do not have lookahead capabilities.\\
\noindent\textbf{Related work:} Various research contributions for the management of \mbox{renewable-powered} \ac{BS} networks that are empowered with computing capabilities exist in literature. The authors of~\cite{xu2016online} incorporate renewables into MEC system and propose an efficient reinforcement \mbox{learning-based} resource management algorithm for handling dynamic workload offloading. Then, in~\cite{edge_controller}, the authors propose a \mbox{controller-based} network architecture for managing \ac{EH} BSs empowered with computation capabilities where on/off switching strategies allow BSs and \acp{VM} to be dynamically switched on/off towards energy savings, over a limited prediction horizon. In~\cite{steering}, the \mbox{green-based} load balancing technique is employed for optimizing \ac{MEC} performance by exploiting the spatial diversity of the available green energy to reshape the network load among the BSs. Here, \mbox{container-based} virtualization was considered. Towards infrastructure sharing, the work of~\cite{gamebasedsharing} employs an infrastructure sharing algorithm towards energy savings. In their work, a \mbox{game-theoretic} framework was proposed in order to enable the MN operators to individually estimate the \mbox{switching-off} probabilities that reduce their expected financial cost. Then, in~\cite{sanguanpuak}, the problem of infrastructure sharing among MN operators is presented as a \mbox{multiple-seller} \mbox{single-buyer} business. In their contribution, each \ac{BS} is utilized by subscribers from other operators and the owner of the BS is considered as a seller of the BS infrastructure while the owners of the subscribers utilizing the BS are considered as buyers. In the presence of multiple seller MN operators, it is assumed that they compete with each other to sell their network infrastructure resources to potential buyers.
However, while performing task offloading, care must be taken in order to avoid large computation delays in overloaded BSs. To avoid these large computation latencies at overloaded BSs, assisted computation must be employed within the MEC paradigm. The works of~\cite{chen2018computation} develops a novel online BS peer offloading framework. Here, a Lyapunov technique is used in order to maximize the \mbox{long-term} system performance while keeping the energy consumption of BSs below individual \mbox{long-term} constraints. In this regard, it is noted that the aforementioned works are not fully offgrid similar to the national park environment that requires no power lines. Also, the joint consideration of infrastructure and resource sharing, and assisted computation offloading for \mbox{energy-constrained} communication sites, within the \ac{MEC} paradigm has not been given enough research attention.\\ 
\begin{figure} [t]
	\centering
	\includegraphics[width = \columnwidth]{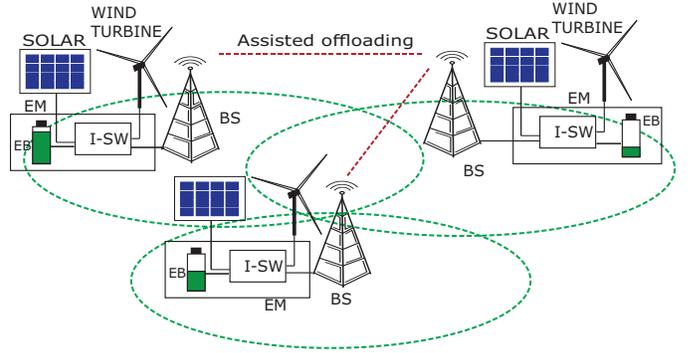}
	\caption{The BS system within the protected area where energy is harvested from the environment.}
	\label{fig:eh_mec}
\end{figure} 
In order to achieve the objective of this paper, the remainder of this article is organized as follows: In Section~\ref{sec:syst}, the system model is described, formulating the optimization problem and solving it in an online fashion via an LLC-based approach. In Section~\ref{sec:res}, some selected results are shown, quantifying the energy savings and delay cost that can be achieved with our framework. Finally, in Section~\ref{sec:concl} some concluding remarks are given.

\section{System Model}
\label{sec:syst}
\begin{figure}[t]
	\centering
	\includegraphics[width = \columnwidth]{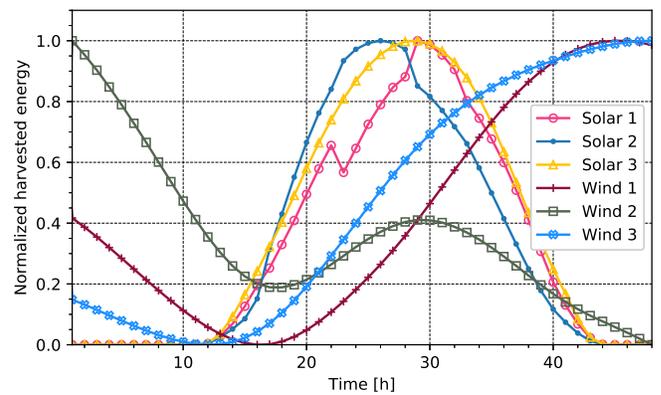}
	\caption{Harvested solar and wind traces~\cite{belgium} used in our simulations.}
	\label{fig:energy_trace}
\end{figure}

We consider $n \in \mathcal{N}$ BSs, indexed by $\mathcal{N} = \{1, \dots, N\}$, deployed in a protected area (national park in our case), each endowed with computing capabilities (BS is \mbox{co-located} with a MEC server that runs $D$ containers), as illustrated in Fig.~\ref{fig:eh_mec} above. Hence, \acp{UE} can offload their computation jobs to corresponding serving BS via wireless communications for processing. In order to preserve the landscape, the deployed BSs have greater level of camouflage. Each communication site is mainly powered by renewable energy harvested from wind and solar radiation, and it is equipped with an \ac{EB} (with a maximum capacity of $B_n^{\rm max}$) for energy storage. There is an Energy Manager (EM) which is responsible for selecting the appropriate energy source to fulfill the \ac{EB}, and also for monitoring the energy level of the \ac{EB}. In addition, the MEC server house a virtualized Access Control Router (ACR) application which acts as an access gateway, responsible for accepting and routing the workloads that are either accepted locally or offloaded to neighboring BSs. For remote clouds or Internet access, a microwave backhaul or a \mbox{multi-hop} wireless backhaul relaying (e.g., integrated  access  and  backhaul (IAB)) is used. 
Moreover, a \mbox{discrete-time} model is considered, whereby the time is discretized as \mbox{$t = 1,2,\dots$} time slots of a fixed duration $\tau = \SI{30}{\minute}$.\\
\noindent\textbf{Communication and computing energy cost}: it is worth noting that the total energy consumed [$\SI{}{\joule}$] per communication site $\theta_n^{\rm site}(t)$ consist of the BS communications, denoted by $\theta_{n}^{\rm bs}(t)$, and computing platform processes, related to computing, caching, and communication, which is denoted by $\theta_n^{\rm mec} (t)$. Thus, the energy consumption model at time slot $t$ is formulated as follows, inspired by~\cite{steering}:
\begin{equation}
	\theta_n^{\rm site}(t) = \theta_{n}^{\rm bs}(t) +  \theta_n^{\rm mec} (t).         
	\label{eq:mecconsupt}
\end{equation}
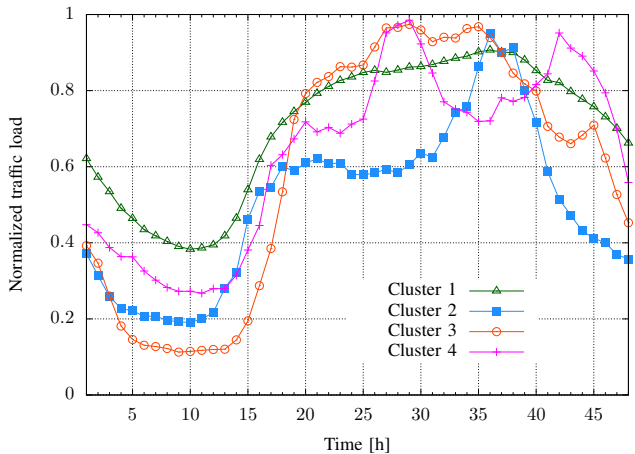
\begin{figure}[t]
	\centering
	\resizebox{\columnwidth}{!}{\input{./traffic_profiles.tex}}
	\caption{Normalized BS traffic loads behavior represented as clusters. The data from~\cite{traces} has been split into four representative clusters.}
	\label{fig:trace_load}
\end{figure}
\noindent\textbf{Energy levels and job flows}: within each BS site there is an intelligent \mbox{electro-mechanical} switch  (I-SW) that aggregates the energy sources to fulfill the \ac{EB} level. In this work, the EBs work as either an energy sources or sinks, depending on traffic load $L_n(t)$ (see Fig.~\ref{fig:trace_load} for traffic load patterns) which consists of \mbox{delay-sensitive} workloads $l_n(t)$. At time slot $t$, the amount of energy required for computation and communication process from the EBs, denoted by $b_n(t) \geq\theta_n^{\rm site}(t)$, is made up of the harvested energy (see Fig.~\ref{fig:energy_trace} above for energy profiles) as follows:
\begin{equation}
  h_n(t) = h_n^c(t) + h_n^o(t) \,,
\end{equation}
where $h_n^c(t)$ is the portion of harvested energy used for charging the battery and $h_n^o(t)$ is the portion of harvested energy immediately used by the site to support its operations. It is worth noting that the actual amount of harvested energy is bounded by the maximum amount of energy that could be harvested from the environment, denoted by $H_n^{\rm max}$, at a certain time slot. Thus, the following hard constraint must hold:
\begin{equation}
  h_n^c(t) + h_n^o(t) \leq H_n^{\rm max} \,, \forall n, \forall t.
\end{equation}
The available EB level, at time slot $t$, located at the BS site (BS $n$) evolves according to the following dynamics:
\begin{equation}
   b_n(t) = \mu_n (b_n(t-1) - \theta_n^{\rm site}(t)) + \alpha_n (h_n^c(t))\,,
   \label{eq:eblevel}
\end{equation} 
where $\mu_n \in (0, 1]$ is a parameter accounting for the \mbox{self-discharging} behavior of the battery, and $\alpha_n \in (0, 1]$ accounts for the losses in the charging process.\\
In the interest of network optimization through infrastructure sharing and resource allocation, over a given horizon, the BSs can perform assisted computation by offloading in full or part the computational tasks, accepted from the connected UEs, to neigboring BSs with sufficient green energy. We define ${w}_{nm} (t)$ (jobs/slot) as the job flow that BS site $n \in \mathcal{N}$ offloads towards BS site $m$ (one among its neighboring BSs) in time slot $t$. Note that ${w}_{nn} (t)$ represents the portion of job flow that is directly collected by MEC server $n$ that is \mbox{co-located} with the BS. At this regard, we also note that the energy required by the BS site $n$ in time slot $t$, obtained from the harvesting systems or EB, is linear and directly proportional to the job flow as follows:
\begin{equation}
 b_n(t) = \eta \left( w_{nn} + \sum_{m \in \mathcal{N} \setminus \{n\}}w_{mn} \right) + \kappa \left( \sum_{m \in \mathcal{N} \setminus \{n\}} w_{nm} \right).
\end{equation}
Note that $\eta > 0$ while $\kappa < 0$ (with $|\eta| > |\kappa|$). Since the admitted computation workloads (jobs) are first buffered at the input buffer at time $t$ and the results are accumulated over a fixed period of time to form a batch at the output buffer. From this, we note that at every time slot the number of jobs
exiting a node cannot exceed the number of jobs entering the
same node. Thus, the following condition holds:
\begin{equation}
    \sum_{m \in \mathcal{N} \setminus \{n\}} (w_{nm}(t) - w_{mn}(t)) \leq w_{nn}(t) \,, \forall t, \forall n .
\end{equation}
\noindent\textbf{Computing resources and load distribution}: the amount of the accepted workload $w_{nn}(t)$ is computed using containers as computing resources within the MEC server. For a fair provisioning of the computing resources, $D(t)$ needs to be obtained first, and then the workload per container $\lambda_d(t)$ is determined. Firstly, each container can only compute an amount of up to $\lambda_{\max}$ (considering that virtualization technologies specify the minimum and maximum amount of load that can be allocated per container) and to meet the latency requirements, $D(t)$ is obtained as: \mbox{$D(t) = \ceil[\big] {(w_{nn}(t)/ \lambda_{\max})}$}, where $\ceil[\big] {\cdot}$ returns the nearest upper integer. Secondly, to distribute the workload among the $D(t)$ containers, a heuristic process splits the computational workload $\lambda_d(t) = \lambda_{\max}$ to the first $D(t)-1$ containers, and the remaining workload $\lambda_{d}(t) = w_{nn}(t) - (D(t)-1)\lambda_{\max}$ to the last one.
\subsection{Optimization}
\label{sub:opt}
Our objective is to improve the overall energy savings
of the communication system through infrastructure sharing principles and resource management procedures, with a QoS guarantee. The power saving modes within each BS site is achieved by enabling assisted computation and also performing the following within the MEC server: autoscaling of containers, contents caching and tuning of the transmission drivers. To achieve our objective, two cost functions are defined, one captures the communication system energy consumption and the other, handles the QoS. This is defined as follows: F1) $\theta_n^{\rm site}(t)$, weighs the energy consumption due to transmission in the BSs and the \mbox{computing-plus-communication} activities in the MEC server. Then, F2) a quadratic term $(w_{nn}(t) - w_{nm}(t))^2$, which accounts for the QoS. At this regard, it is worth noting that F1 tends to push the system towards \mbox{self-sustainability} solutions and F2 favors solutions where the delay sensitive load is either entirely admitted in the \mbox{co-located} MEC server by the router application or offloaded to a neighboring BS that is not \mbox{energy-constrained}. A weight $\Gamma = [0,1]$ is utilized to balance the two objectives F1 and F2. The corresponding (weighted) cost function is defined as:
\begin{equation}
\label{eq:Jfunc_2}
\begin{aligned}
J_n(t) & \stackrel{\Delta}{=} \overline{\Gamma} \, \theta_n^{\rm site}(t)+ \Gamma \, (w_{nn}(t) - w_{nm}(t))^2 \, ,
\end{aligned}
\end{equation}
where $\overline{\Gamma } \stackrel{\Delta}{=} 1 - \Gamma$. Hence, starting from $t = 1$ (i.e., $t = 1,2, \dots, T$) as the current time slot and the finite horizon $T$, the following optimization problem is formulated as:
\begin{eqnarray}
        \label{eq:objt_2}
        \textbf{P1} & : & \min_{h_n^c(t), h_n^o(t),D(t), w_{nm}(t)} \sum_{t=1}^T J_n(t)  \\
        && \hspace{-1.25cm}\mbox{subject to:}\quad \text{Eqn.} (2), (3), (4), (5), (6).\nonumber 
\end{eqnarray}

\noindent To solve {\bf P1} in~\eq{eq:objt_2}, the \ac{LLC} principles~\cite{hayes_2004}, \ac{GP} technique~\cite{geo_prog}, and heuristics, is used towards obtaining the feasible system control inputs.
\subsection{Limited lookahead control approach}
To solve \textbf{P1} (Eq.~(\ref{eq:objt_2})), we are required to have a complete knowledge of the traffic load, harvested energy arrivals, and the workloads accepted by the MEC server. However, such knowledge is not possible to achieve, in fact, we only have knowledge of the exogenous processes in the current, $t$, and past time slots. Thus, we adopt a pragmatic \ac{LLC} approach to solve the problem in an online fashion. Next, we define the system state vector as $q_n(t)$ and the input vector as $\varphi_n(t)$. The system behavior is described by the \mbox{discrete-time} \mbox{state-space} equation, adopting the \ac{LLC} principles~\cite{llcprediction}, as:
\begin{equation}
\ q_n(t + 1) = \phi(q_n(t), \varphi_n(t)) \, , 
\end{equation}
\noindent where  $\phi(\cdot)$ is a behavioral model that captures the relationship between $(q_n(t),\varphi_n(t))$, and the next state $q_n(t + 1)$. Note that this relationship accounts for the amount of energy drained, that harvested $h_n (t)$, which together lead to the next buffer level $b_n(t+1)$. We note that state $q_n(t)$ and $\varphi_n(t)$ are respectively measured and applied at the beginning of the time slot $t$, whereas the offered load $L_n(t)$ and the harvested energy $h_n (t)$ are accumulated during the time slot and their value becomes known only by the end of it. This means that, being at the beginning of time slot $t$, the system state at the next time slot $t+1$ can only be estimated, which is formally written as:
\begin{equation}
       \hat{q}_n(t + 1) = \phi(q_n(t),\varphi_n(t)) \,.
       \label{eq:state_forecast}
\end{equation}
For these estimations, we use the forecast values of load $\hat{L}_n (t)$ and harvested energy $\hat{h}_n (t)$, from the \ac{LSTM} forecasting module (see Section~\ref{sub:predict} for additional details on forecasting). Specifically, for each time slot $t$, problem~\eq{eq:objt_2} is solved obtaining control actions for the prediction horizon $T$. The control action that is applied at time $t$ is $\varphi_n^{*}(t)$, which is the first one in the retrieved control sequence. 
\subsection{Prediction of exogenous processes}\label{sub:predict}
In this section, we describe how the traces used in the simulations were collected and how their predictions were performed.\\\
\noindent\textbf{Traffic load and Harvested energy}: in this work, the amount of harvested energy $h_n(t)$ in time slot $t$ is obtained from \mbox{open-source} solar and wind traces within farm located belgium~\cite{belgium}. As for the traffic load,  open source MN datasets obtained from the \ac{TIM} network (availed through the Big Data Challenge~\cite{bigdata2015tim}) are used to emulate the traffic and computational load. Two exogenous processes are considered in this work: the harvested energy $h_n(t)$ and the BS traffic loads $L_n(t)$. In order to generate the predictions ($\hat{h}_n (t), \hat{L}_n(t)$), the \ac{LSTM} neural networks~\cite{lstmlearn} were adopted. Thus, the \mbox{LSTM-based} predictor has been trained to give an output of the the forecasts for the required number of future time slots $T$. The trained LSTM network consists of an input layer, a single hidden layer consisting of $40$ neurons, for $80$ epochs, for a batch size of $4$; and an output layer. For training and testing purposes, the dataset was split as $70\%$ for training and $30\%$ for testing. As for the performance measure of the model, the \ac{RMSE} is used.\\
\noindent\textbf{Workload (Jobs) flows}: in order to understand the daily traffic load patterns, the clustering algorithm \mbox{X-means}~\cite{pelleg2000x} has been applied to classify the load profiles into several categories. Here, each \ac{BS} $n$ is assumed to have a related load profile ${L}_{n}(t)$ which is picked at random as one of the four clusters in Fig.~\ref{fig:trace_load}. In addition, it is assumed that ${L}_{n}(t)$ consists of $80\%$ delay sensitive workloads $l_{n}(t)$ and the remainder is delay tolerant workloads. Since the amount of the workload to be computed locally, $w_{nn}(t)$, is determined by considering the amount of energy required by BS $n$ at time slot $t$, $b_{n}(t)$, and the next time slot energy to be accumulated through harvesting process, denoted by $b_{n}(t+1)$. Note that the harvested energy is predicted. Thus, $w_{nn}(t) = \frac{b_{n}(t)}{b_n(t+1)} l_n (t)$.
\subsection{The online algorithm}
The online algorithm proceeds as follows, in a distributed manner: Starting from the {\it initial state}, the online algorithm constructs, in a \mbox{breadth-first} fashion, a tree comprising all possible future states up to the prediction depth $T$. 
Then, a search set $\mathcal F$ consisting of the current system state is initialized, and it is accumulated as the algorithm traverse through the tree, accounting for predictions, accumulated workloads at the output buffer, past outputs and controls. The set of states reached at every prediction depth $t+p$ is referred to as $\mathcal F(t+p)$. 
Given $q(t)$, we first estimate the traffic load $\hat{L}_{n}(t+p)$, delay-sensitive jobs $\hat{l}_n(t+p)$, locally acceptable computational load $\hat{w}_{nn}(t + p)$, harvested energy $\hat{h}_n (t+p)$,  any expected acceptable computational jobs from neighbors $\hat{w}_{nm}(t+p)$, and generate the next set of reachable control actions by applying the input workload and energy harvested. 
The energy cost function corresponding to each generated state $q_n(t+p)$ is then computed. Once the prediction horizon  is explored, a sequence of reachable states yielding minimum energy consumption is obtained. The control action $\varphi_n^{*}(t)$ corresponding to $\hat{q}_n(t+p)$ (the first state in this sequence) is provided as input to the system while the rest are discarded. The process is repeated at the beginning of each time slot $t$.
%---------------------------------------------------------------------------------
%       					System Parameters
%-------------------------------------------------------------------------------
\begin{table}[h!]
	\caption{Simulation parameters. Parameters depending on $n$ apply
to all nodes $n \in \mathcal{N}$.}
	\center
	\begin{tabular} {|l| l|l|}
		\hline 
		{\bf Parameter} & {\bf Value} \\ 
		\hline
		$N$ & $20$ \\
		$D$ &  $20$\\
		$B_n^{\rm max}$ & $\SI{100} {\kilo\joule}$\\
		$\mu_n$ & 0.9999 \\
		$\alpha_n$ & 0.900\\
		$\eta$ & $\SI{0.105}{\joule}$/(job/slot) \\
		$\kappa$ & -$\SI{0.035}{\joule}$/(job/slot) \\
		$\lambda_{\rm max}$ & 10 MB\\
		\hline 
	\end{tabular}
	\label{tab_opt}
\end{table}
%----------------------------------------------------------------------------
\section{Performance Evaluations}
\label{sec:res}
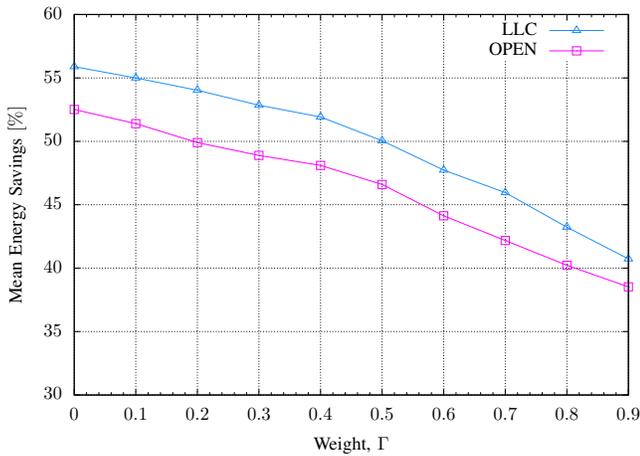
\begin{figure}[t]
	\centering
	\resizebox{\columnwidth}{!}{\input{weight_data.tex}}
	\caption{Overall energy savings versus weighing parameter.}
	\label{fig:qos}
\end{figure}

In this section, some selected numerical results for the
scenario of Section~\ref{sec:syst} are shown. The parameters that were used in the simulations are listed in Table~\ref{tab_opt} above. Our time slot duration $\tau$ is set to $\SI{30}{\minute}$ and the time horizon is set to $T = 3$ time slots. For simulation, Python is used as the programming language.\\
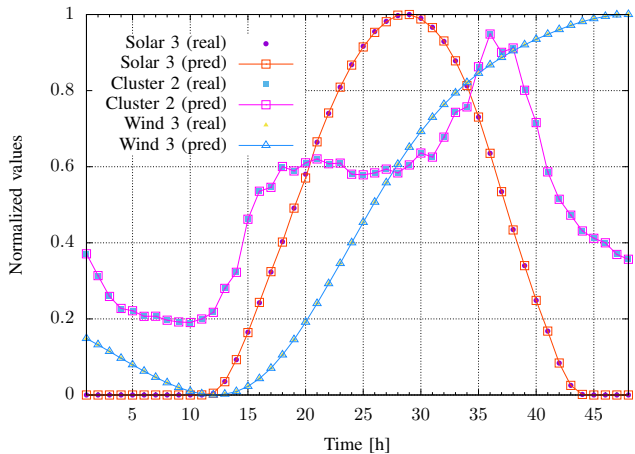
\begin{figure}
	\centering
	\resizebox{\columnwidth}{!}{\input{mixed_data.tex}}
	\caption{Forecast mean value for $L_n(t)$ and $h_n(t)$.}
	\label{fig:mixeddata}
\end{figure}
\textbf{Forecasting}: In Fig.~\ref{fig:mixeddata} above, we show the real and forecasted values for the mobile traffic load (Cluster 2) and harvested energy (Solar 3 and Wind 3) over the time horizon. We show the \mbox{one-step} predictive mean value at each step of the online forecasting routine. The obtained average RMSE for the traffic load and harvested energy processes, both normalized in [0,1] for $T \in \{1,2,3\}$, are $L_{n}(t) = \{0.013, 0.015, 0.020\}$, for solar $h_n(t) = \{0.030, 0.035,0.037\}$, and for wind $h_n(t) = \{0.032, 0.036,0.039\}$. Note that the forecasting for $L_{n}(t)$ are more accurate than those of $h_n(t)$. This is confirmed by comparing the average RMSE. The measured accuracy is deemed good enough for the proposed online optimization.\\ 
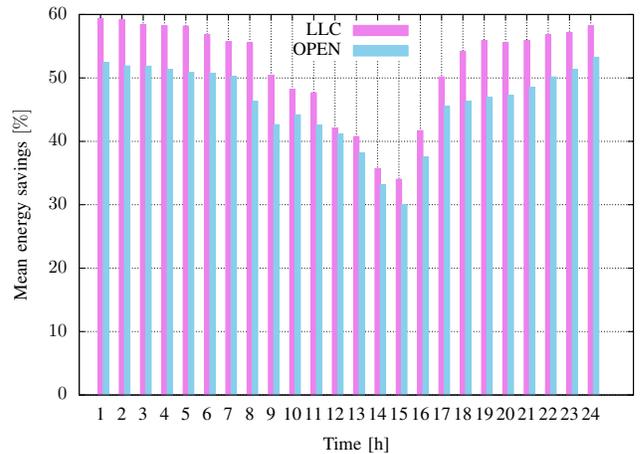
\begin{figure}[t]
	\centering
	\resizebox{\columnwidth}{!}{\input{bssavings.tex}}
	\caption{Mean energy savings.}
	\label{fig:sav}
\end{figure}
\noindent\textbf{Energy consumption}: Our online algorithm (LLC)
is benchmarked with another one, named OPEN (Online BS PEer offloadiNg) from~\cite{chen2018computation}. To address the energy consumption problem, OPEN make use of the Lyapunov technique and solve the problem by using only the current information (no forecasting in this case). Fig.~\ref{fig:sav} above shows the energy savings obtained for one BS with respect to the case where no energy management is performed; i.e., the network is dimensioned for maximum expected capacity. The LLC achieves a energy savings of $52 \%$ and OPEN achieves $46 \%$. The hourly obtained energy savings corresponds to the behavior of national park visitor. In the early morning hours, there are few visitors resulting into higher energy savings ($\SI{6}{\hour} - \SI{8}{\hour}$). During the midday (busy period), the number of visitors arriving increases resulting into reduced amount of energy savings ($\SI{9}{\hour} - \SI{17}{\hour}$). At this time period, assisted computation offloading is more prevalent. After $\SI{17}{\hour}$ there is less activity or no activity thus resulting into an increase in energy savings.\\
\noindent\textbf{Quality of Service (QoS)}: Then, Fig.~\ref{fig:qos} above shows the average energy savings with respect to $\Gamma$ when the BSs ($N = 20$) are running the LLC algorithm in a distributed manner. Here, a \mbox{trade-off} is obtained by either computing all or partial the offload jobs, resulting into either an increase or decrease of energy savings over the protect area. Again, here the energy savings are obtained with respect to the case where all the BSs are dimensioned for maximum capacity. As expected, there is a drop in the energy savings achieved as the value of $\Gamma$ increases, as \ac{QoS} is prioritized. It can be observed that LLC achieves a value of $50 \%$ or above when $\Gamma = [0,0.5]$ and OPEN achieves energy savings of above $50 \%$ when $\Gamma = [0,0.2]$. From this we can observe that LLC outperforms OPEN, thanks to assisted computation offloading and the foresighted optimization which helps to guarantee low latency services.\\

\begin{figure}[t]
	\centering
	\resizebox{\columnwidth}{!}{\input{./mecconsumption.tex}}
	\caption{Energy consumption within the computing platform.}
	\label{fig:mecc}
\end{figure}
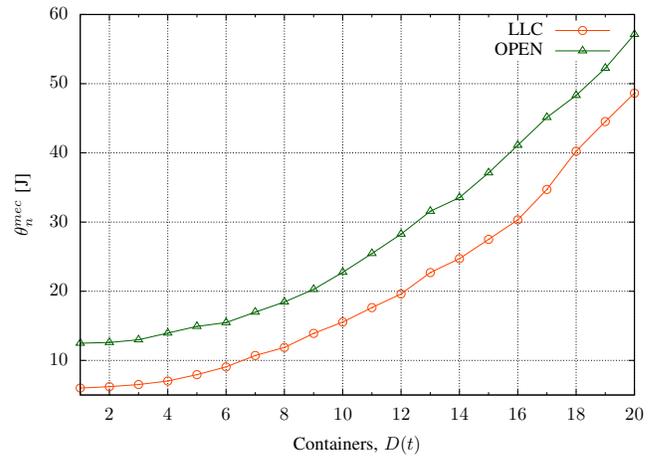

\noindent\textbf{MEC server energy consumption}: Finally, figure~\ref{fig:mecc} above shows the energy drained within the computing platform when the number of active containers is varied. Here, we observe that the lower the number of active containers the lower the energy drained within the server. This correspond to cases where the computational workload $w_{nn}(t)$ is low  (instances of low CPU utilization) or some of the workload have been offloaded to neighboring peers (i.e., assisted computing). From the results, it is observed that LLC consumed less energy when compared with OPEN.
\section{Conclusions}
\label{sec:concl}
The challenge of providing connectivity to protected areas will be one of the pillars for future mobile networks. To address the connectivity issue, this paper formulates an optimization problem to manage a shared base station network (the base station infrastructure and its \mbox{co-located} computing platform that is deployed within a national park. An online algorithm based on infrastructure sharing, assisted computation offloading, forecasting, lookahead principles and heuristics, is proposed with the objective of saving energy within the base stations deployed in a protected area. Numerical results, obtained with \mbox{real-world} energy and traffic load traces, demonstrate that the proposed algorithm achieves mean energy savings of $52\%$ when compared with the $46 \%$ obtained by our benchmark algorithm. The online algorithm achieves energy savings of about $50 \%$ or above when $\Gamma = [0,0.5]$ and the benchmark achieves energy savings of above $50 \%$ when $\Gamma = [0,0.2]$. The energy saving results are obtained with respect to the case where no energy management techniques are applied in the base stations. The obtained results show that there exists a \mbox{trade-off} between energy saving and the QoS. In addition, it is observed that the lower the number of active containers the lower the energy drained within the computing platform, and the increase in the number of active containers results in an increase in the energy being consumed. In this case, assisted computation helps to reduce the energy drained as some of the load is computed by neighboring base stations.

\bibliographystyle{IEEEtran}
\scriptsize
\bibliography{biblio_new}
\end{document}

%% file: traffic_profiles.tex
% GNUPLOT: LaTeX picture with Postscript
\begingroup
  \makeatletter
  \providecommand\color[2][]{%
    \GenericError{(gnuplot) \space\space\space\@spaces}{%
      Package color not loaded in conjunction with
      terminal option `colourtext'%
    }{See the gnuplot documentation for explanation.%
    }{Either use 'blacktext' in gnuplot or load the package
      color.sty in LaTeX.}%
    \renewcommand\color[2][]{}%
  }%
  \providecommand\includegraphics[2][]{%
    \GenericError{(gnuplot) \space\space\space\@spaces}{%
      Package graphicx or graphics not loaded%
    }{See the gnuplot documentation for explanation.%
    }{The gnuplot epslatex terminal needs graphicx.sty or graphics.sty.}%
    \renewcommand\includegraphics[2][]{}%
  }%
  \providecommand\rotatebox[2]{#2}%
  \@ifundefined{ifGPcolor}{%
    \newif\ifGPcolor
    \GPcolortrue
  }{}%
  \@ifundefined{ifGPblacktext}{%
    \newif\ifGPblacktext
    \GPblacktexttrue
  }{}%
  % define a \g@addto@macro without @ in the name:
  \let\gplgaddtomacro\g@addto@macro
  % define empty templates for all commands taking text:
  \gdef\gplbacktext{}%
  \gdef\gplfronttext{}%
  \makeatother
  \ifGPblacktext
    % no textcolor at all
    \def\colorrgb#1{}%
    \def\colorgray#1{}%
  \else
    % gray or color?
    \ifGPcolor
      \def\colorrgb#1{\color[rgb]{#1}}%
      \def\colorgray#1{\color[gray]{#1}}%
      \expandafter\def\csname LTw\endcsname{\color{white}}%
      \expandafter\def\csname LTb\endcsname{\color{black}}%
      \expandafter\def\csname LTa\endcsname{\color{black}}%
      \expandafter\def\csname LT0\endcsname{\color[rgb]{1,0,0}}%
      \expandafter\def\csname LT1\endcsname{\color[rgb]{0,1,0}}%
      \expandafter\def\csname LT2\endcsname{\color[rgb]{0,0,1}}%
      \expandafter\def\csname LT3\endcsname{\color[rgb]{1,0,1}}%
      \expandafter\def\csname LT4\endcsname{\color[rgb]{0,1,1}}%
      \expandafter\def\csname LT5\endcsname{\color[rgb]{1,1,0}}%
      \expandafter\def\csname LT6\endcsname{\color[rgb]{0,0,0}}%
      \expandafter\def\csname LT7\endcsname{\color[rgb]{1,0.3,0}}%
      \expandafter\def\csname LT8\endcsname{\color[rgb]{0.5,0.5,0.5}}%
    \else
      % gray
      \def\colorrgb#1{\color{black}}%
      \def\colorgray#1{\color[gray]{#1}}%
      \expandafter\def\csname LTw\endcsname{\color{white}}%
      \expandafter\def\csname LTb\endcsname{\color{black}}%
      \expandafter\def\csname LTa\endcsname{\color{black}}%
      \expandafter\def\csname LT0\endcsname{\color{black}}%
      \expandafter\def\csname LT1\endcsname{\color{black}}%
      \expandafter\def\csname LT2\endcsname{\color{black}}%
      \expandafter\def\csname LT3\endcsname{\color{black}}%
      \expandafter\def\csname LT4\endcsname{\color{black}}%
      \expandafter\def\csname LT5\endcsname{\color{black}}%
      \expandafter\def\csname LT6\endcsname{\color{black}}%
      \expandafter\def\csname LT7\endcsname{\color{black}}%
      \expandafter\def\csname LT8\endcsname{\color{black}}%
    \fi
  \fi
    \setlength{\unitlength}{0.0500bp}%
    \ifx\gptboxheight\undefined%
      \newlength{\gptboxheight}%
      \newlength{\gptboxwidth}%
      \newsavebox{\gptboxtext}%
    \fi%
    \setlength{\fboxrule}{0.5pt}%
    \setlength{\fboxsep}{1pt}%
\begin{picture}(7200.00,5040.00)%
    \gplgaddtomacro\gplbacktext{%
      \csname LTb\endcsname%%
      \put(814,704){\makebox(0,0)[r]{\strut{}$0$}}%
      \csname LTb\endcsname%%
      \put(814,1527){\makebox(0,0)[r]{\strut{}$0.2$}}%
      \csname LTb\endcsname%%
      \put(814,2350){\makebox(0,0)[r]{\strut{}$0.4$}}%
      \csname LTb\endcsname%%
      \put(814,3173){\makebox(0,0)[r]{\strut{}$0.6$}}%
      \csname LTb\endcsname%%
      \put(814,3996){\makebox(0,0)[r]{\strut{}$0.8$}}%
      \csname LTb\endcsname%%
      \put(814,4819){\makebox(0,0)[r]{\strut{}$1$}}%
      \csname LTb\endcsname%%
      \put(1444,484){\makebox(0,0){\strut{}$5$}}%
      \csname LTb\endcsname%%
      \put(2068,484){\makebox(0,0){\strut{}$10$}}%
      \csname LTb\endcsname%%
      \put(2691,484){\makebox(0,0){\strut{}$15$}}%
      \csname LTb\endcsname%%
      \put(3314,484){\makebox(0,0){\strut{}$20$}}%
      \csname LTb\endcsname%%
      \put(3937,484){\makebox(0,0){\strut{}$25$}}%
      \csname LTb\endcsname%%
      \put(4560,484){\makebox(0,0){\strut{}$30$}}%
      \csname LTb\endcsname%%
      \put(5183,484){\makebox(0,0){\strut{}$35$}}%
      \csname LTb\endcsname%%
      \put(5806,484){\makebox(0,0){\strut{}$40$}}%
      \csname LTb\endcsname%%
      \put(6429,484){\makebox(0,0){\strut{}$45$}}%
    }%
    \gplgaddtomacro\gplfronttext{%
      \csname LTb\endcsname%%
      \put(198,2761){\rotatebox{-270}{\makebox(0,0){\strut{}Normalized traffic load}}}%
      \put(3874,154){\makebox(0,0){\strut{}Time [h]}}%
      \csname LTb\endcsname%%
      \put(4951,1829){\makebox(0,0)[r]{\strut{}Cluster 1}}%
      \csname LTb\endcsname%%
      \put(4951,1609){\makebox(0,0)[r]{\strut{}Cluster 2}}%
      \csname LTb\endcsname%%
      \put(4951,1389){\makebox(0,0)[r]{\strut{}Cluster 3}}%
      \csname LTb\endcsname%%
      \put(4951,1169){\makebox(0,0)[r]{\strut{}Cluster 4}}%
    }%
    \gplbacktext
    \put(0,0){\includegraphics{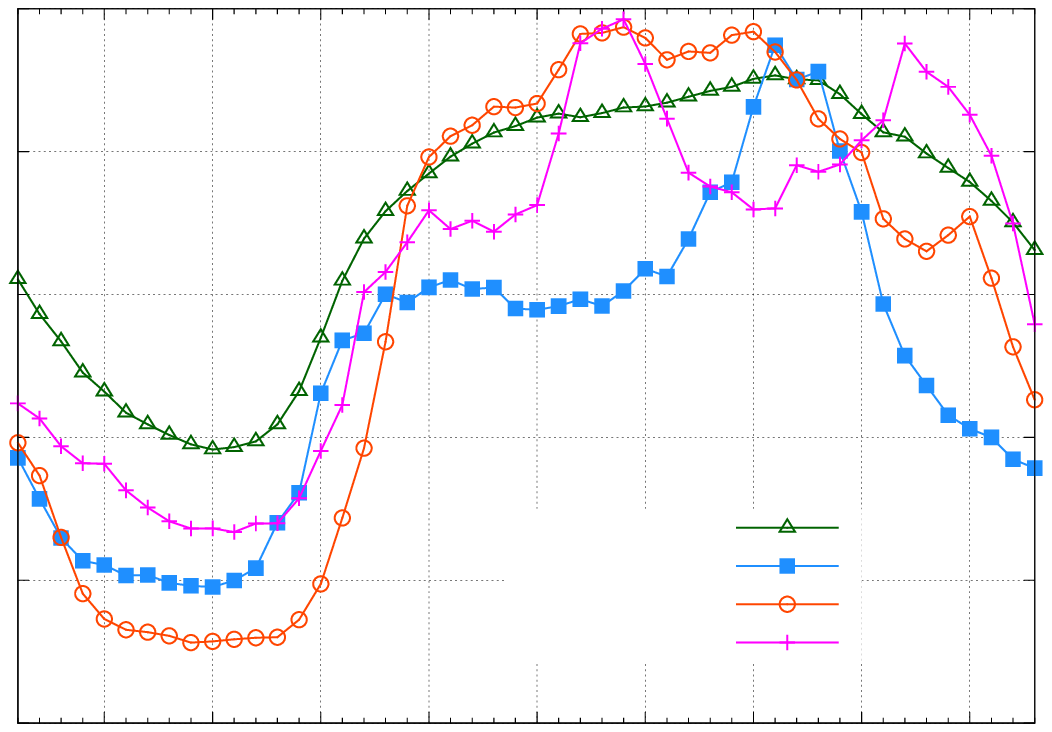}}%
    \gplfronttext
  \end{picture}%
\endgroup

%% file: mixed_data.tex
% GNUPLOT: LaTeX picture with Postscript
\begingroup
  \makeatletter
  \providecommand\color[2][]{%
    \GenericError{(gnuplot) \space\space\space\@spaces}{%
      Package color not loaded in conjunction with
      terminal option `colourtext'%
    }{See the gnuplot documentation for explanation.%
    }{Either use 'blacktext' in gnuplot or load the package
      color.sty in LaTeX.}%
    \renewcommand\color[2][]{}%
  }%
  \providecommand\includegraphics[2][]{%
    \GenericError{(gnuplot) \space\space\space\@spaces}{%
      Package graphicx or graphics not loaded%
    }{See the gnuplot documentation for explanation.%
    }{The gnuplot epslatex terminal needs graphicx.sty or graphics.sty.}%
    \renewcommand\includegraphics[2][]{}%
  }%
  \providecommand\rotatebox[2]{#2}%
  \@ifundefined{ifGPcolor}{%
    \newif\ifGPcolor
    \GPcolortrue
  }{}%
  \@ifundefined{ifGPblacktext}{%
    \newif\ifGPblacktext
    \GPblacktexttrue
  }{}%
  % define a \g@addto@macro without @ in the name:
  \let\gplgaddtomacro\g@addto@macro
  % define empty templates for all commands taking text:
  \gdef\gplbacktext{}%
  \gdef\gplfronttext{}%
  \makeatother
  \ifGPblacktext
    % no textcolor at all
    \def\colorrgb#1{}%
    \def\colorgray#1{}%
  \else
    % gray or color?
    \ifGPcolor
      \def\colorrgb#1{\color[rgb]{#1}}%
      \def\colorgray#1{\color[gray]{#1}}%
      \expandafter\def\csname LTw\endcsname{\color{white}}%
      \expandafter\def\csname LTb\endcsname{\color{black}}%
      \expandafter\def\csname LTa\endcsname{\color{black}}%
      \expandafter\def\csname LT0\endcsname{\color[rgb]{1,0,0}}%
      \expandafter\def\csname LT1\endcsname{\color[rgb]{0,1,0}}%
      \expandafter\def\csname LT2\endcsname{\color[rgb]{0,0,1}}%
      \expandafter\def\csname LT3\endcsname{\color[rgb]{1,0,1}}%
      \expandafter\def\csname LT4\endcsname{\color[rgb]{0,1,1}}%
      \expandafter\def\csname LT5\endcsname{\color[rgb]{1,1,0}}%
      \expandafter\def\csname LT6\endcsname{\color[rgb]{0,0,0}}%
      \expandafter\def\csname LT7\endcsname{\color[rgb]{1,0.3,0}}%
      \expandafter\def\csname LT8\endcsname{\color[rgb]{0.5,0.5,0.5}}%
    \else
      % gray
      \def\colorrgb#1{\color{black}}%
      \def\colorgray#1{\color[gray]{#1}}%
      \expandafter\def\csname LTw\endcsname{\color{white}}%
      \expandafter\def\csname LTb\endcsname{\color{black}}%
      \expandafter\def\csname LTa\endcsname{\color{black}}%
      \expandafter\def\csname LT0\endcsname{\color{black}}%
      \expandafter\def\csname LT1\endcsname{\color{black}}%
      \expandafter\def\csname LT2\endcsname{\color{black}}%
      \expandafter\def\csname LT3\endcsname{\color{black}}%
      \expandafter\def\csname LT4\endcsname{\color{black}}%
      \expandafter\def\csname LT5\endcsname{\color{black}}%
      \expandafter\def\csname LT6\endcsname{\color{black}}%
      \expandafter\def\csname LT7\endcsname{\color{black}}%
      \expandafter\def\csname LT8\endcsname{\color{black}}%
    \fi
  \fi
    \setlength{\unitlength}{0.0500bp}%
    \ifx\gptboxheight\undefined%
      \newlength{\gptboxheight}%
      \newlength{\gptboxwidth}%
      \newsavebox{\gptboxtext}%
    \fi%
    \setlength{\fboxrule}{0.5pt}%
    \setlength{\fboxsep}{1pt}%
\begin{picture}(7200.00,5040.00)%
    \gplgaddtomacro\gplbacktext{%
      \csname LTb\endcsname%%
      \put(814,704){\makebox(0,0)[r]{\strut{}$0$}}%
      \csname LTb\endcsname%%
      \put(814,1527){\makebox(0,0)[r]{\strut{}$0.2$}}%
      \csname LTb\endcsname%%
      \put(814,2350){\makebox(0,0)[r]{\strut{}$0.4$}}%
      \csname LTb\endcsname%%
      \put(814,3173){\makebox(0,0)[r]{\strut{}$0.6$}}%
      \csname LTb\endcsname%%
      \put(814,3996){\makebox(0,0)[r]{\strut{}$0.8$}}%
      \csname LTb\endcsname%%
      \put(814,4819){\makebox(0,0)[r]{\strut{}$1$}}%
      \csname LTb\endcsname%%
      \put(1444,484){\makebox(0,0){\strut{}$5$}}%
      \csname LTb\endcsname%%
      \put(2068,484){\makebox(0,0){\strut{}$10$}}%
      \csname LTb\endcsname%%
      \put(2691,484){\makebox(0,0){\strut{}$15$}}%
      \csname LTb\endcsname%%
      \put(3314,484){\makebox(0,0){\strut{}$20$}}%
      \csname LTb\endcsname%%
      \put(3937,484){\makebox(0,0){\strut{}$25$}}%
      \csname LTb\endcsname%%
      \put(4560,484){\makebox(0,0){\strut{}$30$}}%
      \csname LTb\endcsname%%
      \put(5183,484){\makebox(0,0){\strut{}$35$}}%
      \csname LTb\endcsname%%
      \put(5806,484){\makebox(0,0){\strut{}$40$}}%
      \csname LTb\endcsname%%
      \put(6429,484){\makebox(0,0){\strut{}$45$}}%
    }%
    \gplgaddtomacro\gplfronttext{%
      \csname LTb\endcsname%%
      \put(198,2761){\rotatebox{-270}{\makebox(0,0){\strut{}Normalized values}}}%
      \put(3874,154){\makebox(0,0){\strut{}Time [h]}}%
      \csname LTb\endcsname%%
      \put(2459,4503){\makebox(0,0)[r]{\strut{}Solar 3 (real)}}%
      \csname LTb\endcsname%%
      \put(2459,4283){\makebox(0,0)[r]{\strut{}Solar 3 (pred)}}%
      \csname LTb\endcsname%%
      \put(2459,4063){\makebox(0,0)[r]{\strut{}Cluster 2 (real)}}%
      \csname LTb\endcsname%%
      \put(2459,3843){\makebox(0,0)[r]{\strut{}Cluster 2 (pred)}}%
      \csname LTb\endcsname%%
      \put(2459,3623){\makebox(0,0)[r]{\strut{} Wind 3 (real)}}%
      \csname LTb\endcsname%%
      \put(2459,3403){\makebox(0,0)[r]{\strut{}Wind 3 (pred)}}%
    }%
    \gplbacktext
    \put(0,0){\includegraphics{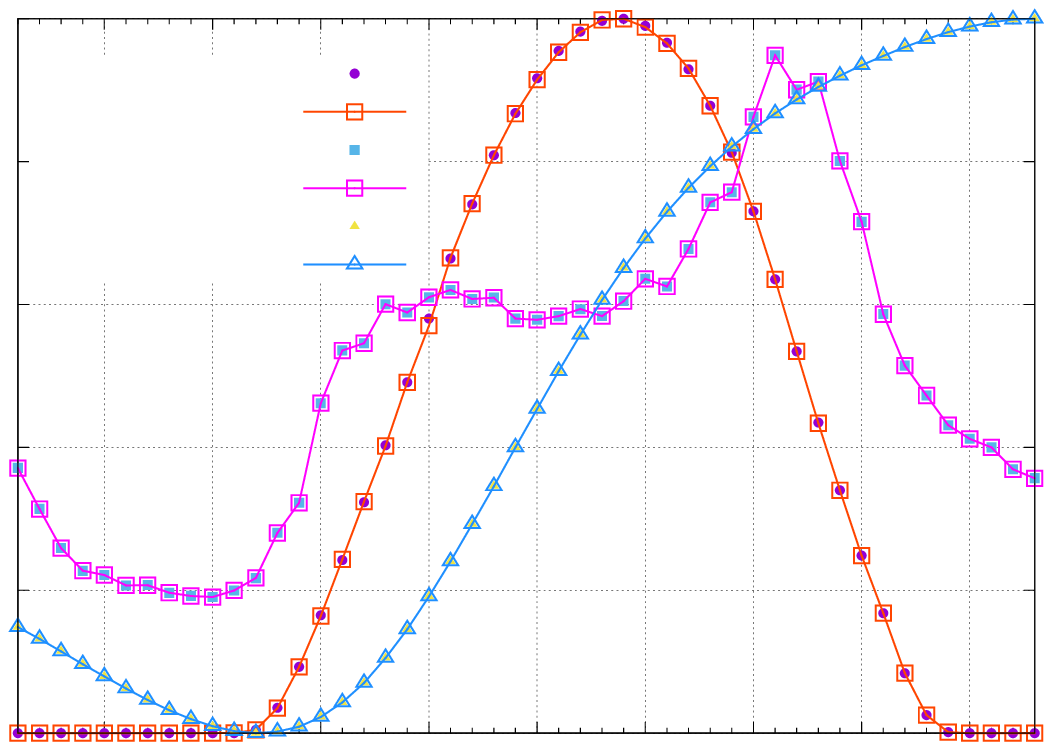}}%
    \gplfronttext
  \end{picture}%
\endgroup

%% file: bssavings.tex
% GNUPLOT: LaTeX picture with Postscript
\begingroup
  \makeatletter
  \providecommand\color[2][]{%
    \GenericError{(gnuplot) \space\space\space\@spaces}{%
      Package color not loaded in conjunction with
      terminal option `colourtext'%
    }{See the gnuplot documentation for explanation.%
    }{Either use 'blacktext' in gnuplot or load the package
      color.sty in LaTeX.}%
    \renewcommand\color[2][]{}%
  }%
  \providecommand\includegraphics[2][]{%
    \GenericError{(gnuplot) \space\space\space\@spaces}{%
      Package graphicx or graphics not loaded%
    }{See the gnuplot documentation for explanation.%
    }{The gnuplot epslatex terminal needs graphicx.sty or graphics.sty.}%
    \renewcommand\includegraphics[2][]{}%
  }%
  \providecommand\rotatebox[2]{#2}%
  \@ifundefined{ifGPcolor}{%
    \newif\ifGPcolor
    \GPcolortrue
  }{}%
  \@ifundefined{ifGPblacktext}{%
    \newif\ifGPblacktext
    \GPblacktexttrue
  }{}%
  % define a \g@addto@macro without @ in the name:
  \let\gplgaddtomacro\g@addto@macro
  % define empty templates for all commands taking text:
  \gdef\gplbacktext{}%
  \gdef\gplfronttext{}%
  \makeatother
  \ifGPblacktext
    % no textcolor at all
    \def\colorrgb#1{}%
    \def\colorgray#1{}%
  \else
    % gray or color?
    \ifGPcolor
      \def\colorrgb#1{\color[rgb]{#1}}%
      \def\colorgray#1{\color[gray]{#1}}%
      \expandafter\def\csname LTw\endcsname{\color{white}}%
      \expandafter\def\csname LTb\endcsname{\color{black}}%
      \expandafter\def\csname LTa\endcsname{\color{black}}%
      \expandafter\def\csname LT0\endcsname{\color[rgb]{1,0,0}}%
      \expandafter\def\csname LT1\endcsname{\color[rgb]{0,1,0}}%
      \expandafter\def\csname LT2\endcsname{\color[rgb]{0,0,1}}%
      \expandafter\def\csname LT3\endcsname{\color[rgb]{1,0,1}}%
      \expandafter\def\csname LT4\endcsname{\color[rgb]{0,1,1}}%
      \expandafter\def\csname LT5\endcsname{\color[rgb]{1,1,0}}%
      \expandafter\def\csname LT6\endcsname{\color[rgb]{0,0,0}}%
      \expandafter\def\csname LT7\endcsname{\color[rgb]{1,0.3,0}}%
      \expandafter\def\csname LT8\endcsname{\color[rgb]{0.5,0.5,0.5}}%
    \else
      % gray
      \def\colorrgb#1{\color{black}}%
      \def\colorgray#1{\color[gray]{#1}}%
      \expandafter\def\csname LTw\endcsname{\color{white}}%
      \expandafter\def\csname LTb\endcsname{\color{black}}%
      \expandafter\def\csname LTa\endcsname{\color{black}}%
      \expandafter\def\csname LT0\endcsname{\color{black}}%
      \expandafter\def\csname LT1\endcsname{\color{black}}%
      \expandafter\def\csname LT2\endcsname{\color{black}}%
      \expandafter\def\csname LT3\endcsname{\color{black}}%
      \expandafter\def\csname LT4\endcsname{\color{black}}%
      \expandafter\def\csname LT5\endcsname{\color{black}}%
      \expandafter\def\csname LT6\endcsname{\color{black}}%
      \expandafter\def\csname LT7\endcsname{\color{black}}%
      \expandafter\def\csname LT8\endcsname{\color{black}}%
    \fi
  \fi
    \setlength{\unitlength}{0.0500bp}%
    \ifx\gptboxheight\undefined%
      \newlength{\gptboxheight}%
      \newlength{\gptboxwidth}%
      \newsavebox{\gptboxtext}%
    \fi%
    \setlength{\fboxrule}{0.5pt}%
    \setlength{\fboxsep}{1pt}%
\begin{picture}(7200.00,5040.00)%
    \gplgaddtomacro\gplbacktext{%
      \csname LTb\endcsname%%
      \put(682,704){\makebox(0,0)[r]{\strut{}$0$}}%
      \csname LTb\endcsname%%
      \put(682,1390){\makebox(0,0)[r]{\strut{}$10$}}%
      \csname LTb\endcsname%%
      \put(682,2076){\makebox(0,0)[r]{\strut{}$20$}}%
      \csname LTb\endcsname%%
      \put(682,2762){\makebox(0,0)[r]{\strut{}$30$}}%
      \csname LTb\endcsname%%
      \put(682,3447){\makebox(0,0)[r]{\strut{}$40$}}%
      \csname LTb\endcsname%%
      \put(682,4133){\makebox(0,0)[r]{\strut{}$50$}}%
      \csname LTb\endcsname%%
      \put(682,4819){\makebox(0,0)[r]{\strut{}$60$}}%
      \csname LTb\endcsname%%
      \put(1044,484){\makebox(0,0){\strut{}1}}%
      \csname LTb\endcsname%%
      \put(1275,484){\makebox(0,0){\strut{}2}}%
      \csname LTb\endcsname%%
      \put(1505,484){\makebox(0,0){\strut{}3}}%
      \csname LTb\endcsname%%
      \put(1735,484){\makebox(0,0){\strut{}4}}%
      \csname LTb\endcsname%%
      \put(1966,484){\makebox(0,0){\strut{}5}}%
      \csname LTb\endcsname%%
      \put(2196,484){\makebox(0,0){\strut{}6}}%
      \csname LTb\endcsname%%
      \put(2426,484){\makebox(0,0){\strut{}7}}%
      \csname LTb\endcsname%%
      \put(2657,484){\makebox(0,0){\strut{}8}}%
      \csname LTb\endcsname%%
      \put(2887,484){\makebox(0,0){\strut{}9}}%
      \csname LTb\endcsname%%
      \put(3117,484){\makebox(0,0){\strut{}10}}%
      \csname LTb\endcsname%%
      \put(3348,484){\makebox(0,0){\strut{}11}}%
      \csname LTb\endcsname%%
      \put(3578,484){\makebox(0,0){\strut{}12}}%
      \csname LTb\endcsname%%
      \put(3809,484){\makebox(0,0){\strut{}13}}%
      \csname LTb\endcsname%%
      \put(4039,484){\makebox(0,0){\strut{}14}}%
      \csname LTb\endcsname%%
      \put(4269,484){\makebox(0,0){\strut{}15}}%
      \csname LTb\endcsname%%
      \put(4500,484){\makebox(0,0){\strut{}16}}%
      \csname LTb\endcsname%%
      \put(4730,484){\makebox(0,0){\strut{}17}}%
      \csname LTb\endcsname%%
      \put(4960,484){\makebox(0,0){\strut{}18}}%
      \csname LTb\endcsname%%
      \put(5191,484){\makebox(0,0){\strut{}19}}%
      \csname LTb\endcsname%%
      \put(5421,484){\makebox(0,0){\strut{}20}}%
      \csname LTb\endcsname%%
      \put(5651,484){\makebox(0,0){\strut{}21}}%
      \csname LTb\endcsname%%
      \put(5882,484){\makebox(0,0){\strut{}22}}%
      \csname LTb\endcsname%%
      \put(6112,484){\makebox(0,0){\strut{}23}}%
      \csname LTb\endcsname%%
      \put(6342,484){\makebox(0,0){\strut{}24}}%
    }%
    \gplgaddtomacro\gplfronttext{%
      \csname LTb\endcsname%%
      \put(198,2761){\rotatebox{-270}{\makebox(0,0){\strut{}Mean energy savings $[\%]$}}}%
      \put(3808,154){\makebox(0,0){\strut{}Time [h]}}%
      \csname LTb\endcsname%%
      \put(3645,4646){\makebox(0,0)[r]{\strut{}LLC}}%
      \csname LTb\endcsname%%
      \put(3645,4426){\makebox(0,0)[r]{\strut{}OPEN}}%
    }%
    \gplbacktext
    \put(0,0){\includegraphics{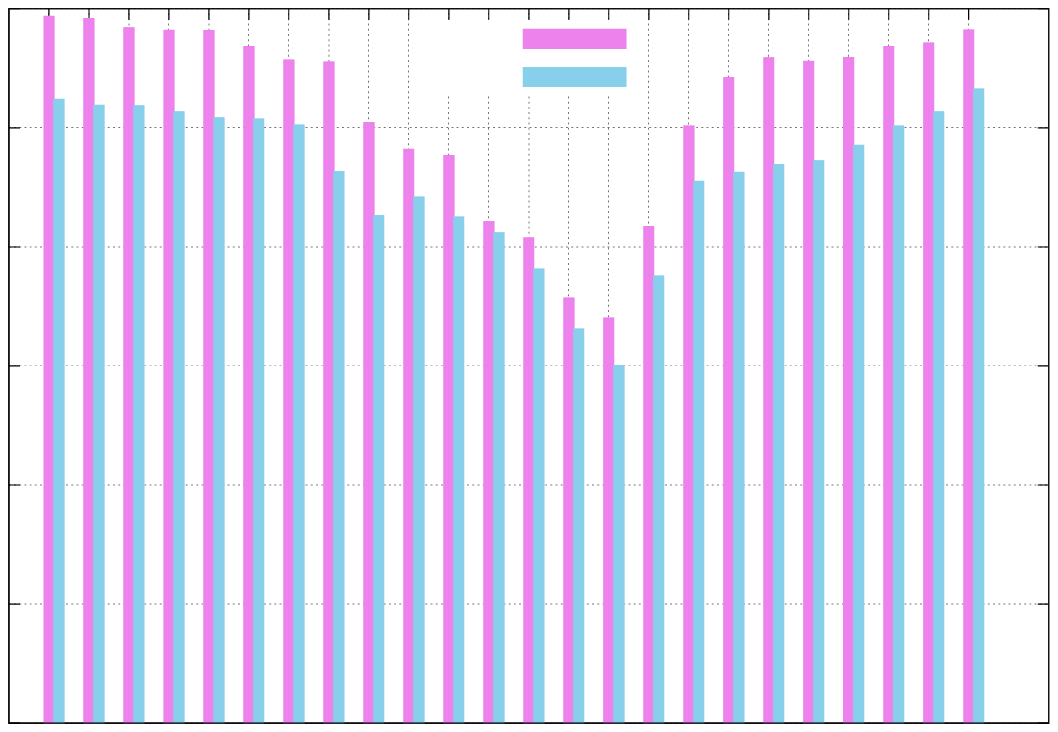}}%
    \gplfronttext
  \end{picture}%
\endgroup

%% file: weight_data.tex
% GNUPLOT: LaTeX picture with Postscript
\begingroup
  \makeatletter
  \providecommand\color[2][]{%
    \GenericError{(gnuplot) \space\space\space\@spaces}{%
      Package color not loaded in conjunction with
      terminal option `colourtext'%
    }{See the gnuplot documentation for explanation.%
    }{Either use 'blacktext' in gnuplot or load the package
      color.sty in LaTeX.}%
    \renewcommand\color[2][]{}%
  }%
  \providecommand\includegraphics[2][]{%
    \GenericError{(gnuplot) \space\space\space\@spaces}{%
      Package graphicx or graphics not loaded%
    }{See the gnuplot documentation for explanation.%
    }{The gnuplot epslatex terminal needs graphicx.sty or graphics.sty.}%
    \renewcommand\includegraphics[2][]{}%
  }%
  \providecommand\rotatebox[2]{#2}%
  \@ifundefined{ifGPcolor}{%
    \newif\ifGPcolor
    \GPcolortrue
  }{}%
  \@ifundefined{ifGPblacktext}{%
    \newif\ifGPblacktext
    \GPblacktexttrue
  }{}%
  % define a \g@addto@macro without @ in the name:
  \let\gplgaddtomacro\g@addto@macro
  % define empty templates for all commands taking text:
  \gdef\gplbacktext{}%
  \gdef\gplfronttext{}%
  \makeatother
  \ifGPblacktext
    % no textcolor at all
    \def\colorrgb#1{}%
    \def\colorgray#1{}%
  \else
    % gray or color?
    \ifGPcolor
      \def\colorrgb#1{\color[rgb]{#1}}%
      \def\colorgray#1{\color[gray]{#1}}%
      \expandafter\def\csname LTw\endcsname{\color{white}}%
      \expandafter\def\csname LTb\endcsname{\color{black}}%
      \expandafter\def\csname LTa\endcsname{\color{black}}%
      \expandafter\def\csname LT0\endcsname{\color[rgb]{1,0,0}}%
      \expandafter\def\csname LT1\endcsname{\color[rgb]{0,1,0}}%
      \expandafter\def\csname LT2\endcsname{\color[rgb]{0,0,1}}%
      \expandafter\def\csname LT3\endcsname{\color[rgb]{1,0,1}}%
      \expandafter\def\csname LT4\endcsname{\color[rgb]{0,1,1}}%
      \expandafter\def\csname LT5\endcsname{\color[rgb]{1,1,0}}%
      \expandafter\def\csname LT6\endcsname{\color[rgb]{0,0,0}}%
      \expandafter\def\csname LT7\endcsname{\color[rgb]{1,0.3,0}}%
      \expandafter\def\csname LT8\endcsname{\color[rgb]{0.5,0.5,0.5}}%
    \else
      % gray
      \def\colorrgb#1{\color{black}}%
      \def\colorgray#1{\color[gray]{#1}}%
      \expandafter\def\csname LTw\endcsname{\color{white}}%
      \expandafter\def\csname LTb\endcsname{\color{black}}%
      \expandafter\def\csname LTa\endcsname{\color{black}}%
      \expandafter\def\csname LT0\endcsname{\color{black}}%
      \expandafter\def\csname LT1\endcsname{\color{black}}%
      \expandafter\def\csname LT2\endcsname{\color{black}}%
      \expandafter\def\csname LT3\endcsname{\color{black}}%
      \expandafter\def\csname LT4\endcsname{\color{black}}%
      \expandafter\def\csname LT5\endcsname{\color{black}}%
      \expandafter\def\csname LT6\endcsname{\color{black}}%
      \expandafter\def\csname LT7\endcsname{\color{black}}%
      \expandafter\def\csname LT8\endcsname{\color{black}}%
    \fi
  \fi
    \setlength{\unitlength}{0.0500bp}%
    \ifx\gptboxheight\undefined%
      \newlength{\gptboxheight}%
      \newlength{\gptboxwidth}%
      \newsavebox{\gptboxtext}%
    \fi%
    \setlength{\fboxrule}{0.5pt}%
    \setlength{\fboxsep}{1pt}%
\begin{picture}(7200.00,5040.00)%
    \gplgaddtomacro\gplbacktext{%
      \csname LTb\endcsname%%
      \put(682,704){\makebox(0,0)[r]{\strut{}$30$}}%
      \csname LTb\endcsname%%
      \put(682,1390){\makebox(0,0)[r]{\strut{}$35$}}%
      \csname LTb\endcsname%%
      \put(682,2076){\makebox(0,0)[r]{\strut{}$40$}}%
      \csname LTb\endcsname%%
      \put(682,2762){\makebox(0,0)[r]{\strut{}$45$}}%
      \csname LTb\endcsname%%
      \put(682,3447){\makebox(0,0)[r]{\strut{}$50$}}%
      \csname LTb\endcsname%%
      \put(682,4133){\makebox(0,0)[r]{\strut{}$55$}}%
      \csname LTb\endcsname%%
      \put(682,4819){\makebox(0,0)[r]{\strut{}$60$}}%
      \csname LTb\endcsname%%
      \put(814,484){\makebox(0,0){\strut{}$0$}}%
      \csname LTb\endcsname%%
      \put(1479,484){\makebox(0,0){\strut{}$0.1$}}%
      \csname LTb\endcsname%%
      \put(2145,484){\makebox(0,0){\strut{}$0.2$}}%
      \csname LTb\endcsname%%
      \put(2810,484){\makebox(0,0){\strut{}$0.3$}}%
      \csname LTb\endcsname%%
      \put(3476,484){\makebox(0,0){\strut{}$0.4$}}%
      \csname LTb\endcsname%%
      \put(4141,484){\makebox(0,0){\strut{}$0.5$}}%
      \csname LTb\endcsname%%
      \put(4807,484){\makebox(0,0){\strut{}$0.6$}}%
      \csname LTb\endcsname%%
      \put(5472,484){\makebox(0,0){\strut{}$0.7$}}%
      \csname LTb\endcsname%%
      \put(6138,484){\makebox(0,0){\strut{}$0.8$}}%
      \csname LTb\endcsname%%
      \put(6803,484){\makebox(0,0){\strut{}$0.9$}}%
    }%
    \gplgaddtomacro\gplfronttext{%
      \csname LTb\endcsname%%
      \put(198,2761){\rotatebox{-270}{\makebox(0,0){\strut{}Mean Energy Savings $[\%]$}}}%
      \put(3808,154){\makebox(0,0){\strut{}Weight, $\Gamma$}}%
      \csname LTb\endcsname%%
      \put(5816,4646){\makebox(0,0)[r]{\strut{}LLC}}%
      \csname LTb\endcsname%%
      \put(5816,4426){\makebox(0,0)[r]{\strut{}OPEN}}%
    }%
    \gplbacktext
    \put(0,0){\includegraphics{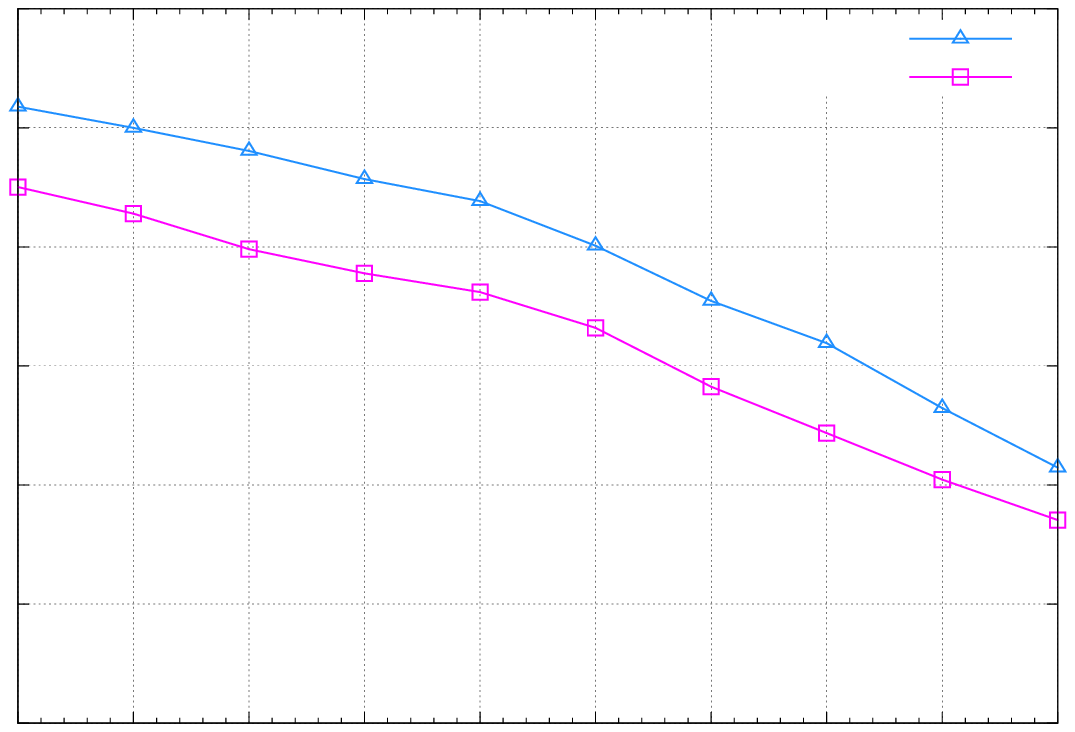}}%
    \gplfronttext
  \end{picture}%
\endgroup

%% file: mecconsumption.tex
% GNUPLOT: LaTeX picture with Postscript
\begingroup
  \makeatletter
  \providecommand\color[2][]{%
    \GenericError{(gnuplot) \space\space\space\@spaces}{%
      Package color not loaded in conjunction with
      terminal option `colourtext'%
    }{See the gnuplot documentation for explanation.%
    }{Either use 'blacktext' in gnuplot or load the package
      color.sty in LaTeX.}%
    \renewcommand\color[2][]{}%
  }%
  \providecommand\includegraphics[2][]{%
    \GenericError{(gnuplot) \space\space\space\@spaces}{%
      Package graphicx or graphics not loaded%
    }{See the gnuplot documentation for explanation.%
    }{The gnuplot epslatex terminal needs graphicx.sty or graphics.sty.}%
    \renewcommand\includegraphics[2][]{}%
  }%
  \providecommand\rotatebox[2]{#2}%
  \@ifundefined{ifGPcolor}{%
    \newif\ifGPcolor
    \GPcolortrue
  }{}%
  \@ifundefined{ifGPblacktext}{%
    \newif\ifGPblacktext
    \GPblacktexttrue
  }{}%
  % define a \g@addto@macro without @ in the name:
  \let\gplgaddtomacro\g@addto@macro
  % define empty templates for all commands taking text:
  \gdef\gplbacktext{}%
  \gdef\gplfronttext{}%
  \makeatother
  \ifGPblacktext
    % no textcolor at all
    \def\colorrgb#1{}%
    \def\colorgray#1{}%
  \else
    % gray or color?
    \ifGPcolor
      \def\colorrgb#1{\color[rgb]{#1}}%
      \def\colorgray#1{\color[gray]{#1}}%
      \expandafter\def\csname LTw\endcsname{\color{white}}%
      \expandafter\def\csname LTb\endcsname{\color{black}}%
      \expandafter\def\csname LTa\endcsname{\color{black}}%
      \expandafter\def\csname LT0\endcsname{\color[rgb]{1,0,0}}%
      \expandafter\def\csname LT1\endcsname{\color[rgb]{0,1,0}}%
      \expandafter\def\csname LT2\endcsname{\color[rgb]{0,0,1}}%
      \expandafter\def\csname LT3\endcsname{\color[rgb]{1,0,1}}%
      \expandafter\def\csname LT4\endcsname{\color[rgb]{0,1,1}}%
      \expandafter\def\csname LT5\endcsname{\color[rgb]{1,1,0}}%
      \expandafter\def\csname LT6\endcsname{\color[rgb]{0,0,0}}%
      \expandafter\def\csname LT7\endcsname{\color[rgb]{1,0.3,0}}%
      \expandafter\def\csname LT8\endcsname{\color[rgb]{0.5,0.5,0.5}}%
    \else
      % gray
      \def\colorrgb#1{\color{black}}%
      \def\colorgray#1{\color[gray]{#1}}%
      \expandafter\def\csname LTw\endcsname{\color{white}}%
      \expandafter\def\csname LTb\endcsname{\color{black}}%
      \expandafter\def\csname LTa\endcsname{\color{black}}%
      \expandafter\def\csname LT0\endcsname{\color{black}}%
      \expandafter\def\csname LT1\endcsname{\color{black}}%
      \expandafter\def\csname LT2\endcsname{\color{black}}%
      \expandafter\def\csname LT3\endcsname{\color{black}}%
      \expandafter\def\csname LT4\endcsname{\color{black}}%
      \expandafter\def\csname LT5\endcsname{\color{black}}%
      \expandafter\def\csname LT6\endcsname{\color{black}}%
      \expandafter\def\csname LT7\endcsname{\color{black}}%
      \expandafter\def\csname LT8\endcsname{\color{black}}%
    \fi
  \fi
    \setlength{\unitlength}{0.0500bp}%
    \ifx\gptboxheight\undefined%
      \newlength{\gptboxheight}%
      \newlength{\gptboxwidth}%
      \newsavebox{\gptboxtext}%
    \fi%
    \setlength{\fboxrule}{0.5pt}%
    \setlength{\fboxsep}{1pt}%
\begin{picture}(7200.00,5040.00)%
    \gplgaddtomacro\gplbacktext{%
      \csname LTb\endcsname%%
      \put(682,1078){\makebox(0,0)[r]{\strut{}$10$}}%
      \csname LTb\endcsname%%
      \put(682,1826){\makebox(0,0)[r]{\strut{}$20$}}%
      \csname LTb\endcsname%%
      \put(682,2574){\makebox(0,0)[r]{\strut{}$30$}}%
      \csname LTb\endcsname%%
      \put(682,3323){\makebox(0,0)[r]{\strut{}$40$}}%
      \csname LTb\endcsname%%
      \put(682,4071){\makebox(0,0)[r]{\strut{}$50$}}%
      \csname LTb\endcsname%%
      \put(682,4819){\makebox(0,0)[r]{\strut{}$60$}}%
      \csname LTb\endcsname%%
      \put(1129,484){\makebox(0,0){\strut{}$2$}}%
      \csname LTb\endcsname%%
      \put(1760,484){\makebox(0,0){\strut{}$4$}}%
      \csname LTb\endcsname%%
      \put(2390,484){\makebox(0,0){\strut{}$6$}}%
      \csname LTb\endcsname%%
      \put(3020,484){\makebox(0,0){\strut{}$8$}}%
      \csname LTb\endcsname%%
      \put(3651,484){\makebox(0,0){\strut{}$10$}}%
      \csname LTb\endcsname%%
      \put(4281,484){\makebox(0,0){\strut{}$12$}}%
      \csname LTb\endcsname%%
      \put(4912,484){\makebox(0,0){\strut{}$14$}}%
      \csname LTb\endcsname%%
      \put(5542,484){\makebox(0,0){\strut{}$16$}}%
      \csname LTb\endcsname%%
      \put(6173,484){\makebox(0,0){\strut{}$18$}}%
      \csname LTb\endcsname%%
      \put(6803,484){\makebox(0,0){\strut{}$20$}}%
    }%
    \gplgaddtomacro\gplfronttext{%
      \csname LTb\endcsname%%
      \put(198,2761){\rotatebox{-270}{\makebox(0,0){\strut{}$\theta_{n}^{mec}$ [J]}}}%
      \put(3808,154){\makebox(0,0){\strut{}Containers, $D(t)$}}%
      \csname LTb\endcsname%%
      \put(5816,4646){\makebox(0,0)[r]{\strut{}LLC}}%
      \csname LTb\endcsname%%
      \put(5816,4426){\makebox(0,0)[r]{\strut{}OPEN}}%
    }%
    \gplbacktext
    \put(0,0){\includegraphics{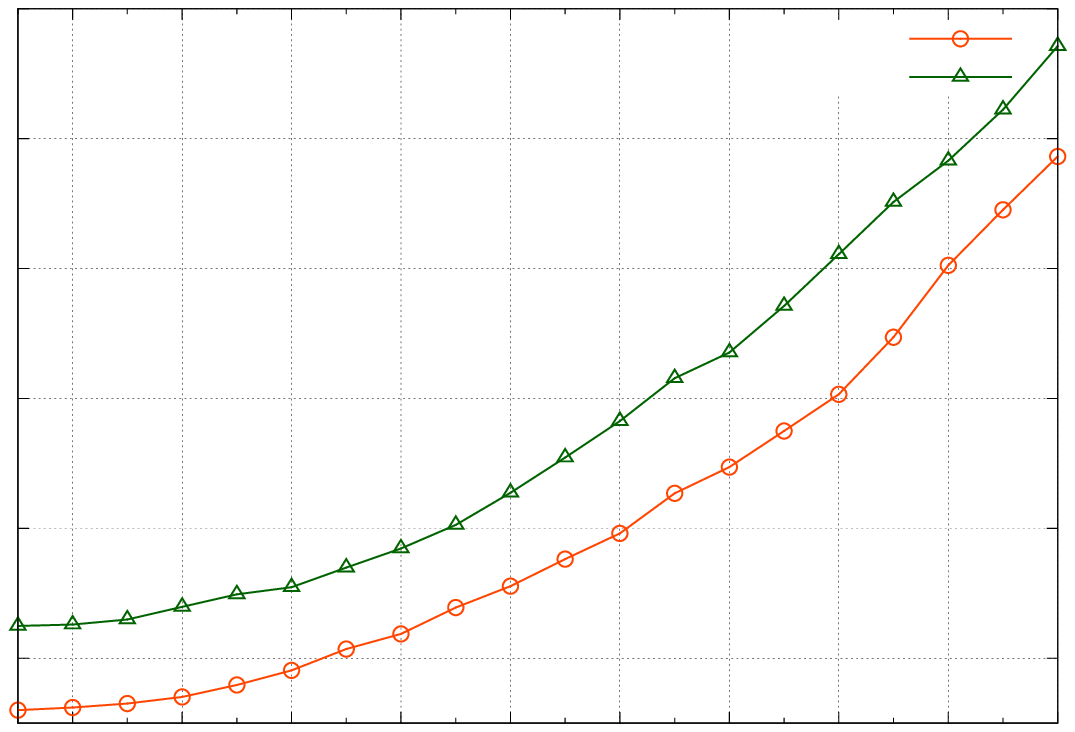}}%
    \gplfronttext
  \end{picture}%
\endgroup